\documentclass[12pt,titlepage,letterpaper]{utarticle}

\usepackage{amsmath}
\usepackage{mathrsfs}
\usepackage{amsfonts}
\usepackage{amssymb}
\usepackage{amsthm}
\usepackage{mathtools}
\usepackage{mathbbol}
\usepackage{graphicx}
\usepackage{color}
\usepackage{xparse}
\usepackage{tikz}
\usepackage{tocloft}
\usepackage[titletoc,title]{appendix}
\usepackage[nosort]{cite}
\usepackage{hyperref}
\usepackage{longtable}
\usepackage{comment}

\usepackage{caption}

\usetikzlibrary{cd}

\definecolor{lightmauve}{RGB}{255,187,255}
\definecolor{lightblue}{RGB}{238,238,255}
\definecolor{lightred}{RGB}{255,238,238}
\definecolor{midgreen}{RGB}{15,170,15}

\definecolor{aqua}{rgb}{0, 1.0, 1.0}
\definecolor{fuschia}{rgb}{1.0, 0, 1.0}
\definecolor{gray}{rgb}{0.502, 0.502, 0.502}
\definecolor{lime}{rgb}{0, 1.0, 0}
\definecolor{maroon}{rgb}{0.502, 0, 0}
\definecolor{navy}{rgb}{0, 0, 0.502}
\definecolor{olive}{rgb}{0.502, 0.502, 0}
\definecolor{purple}{rgb}{0.502, 0, 0.502}
\definecolor{silver}{rgb}{0.753, 0.753, 0.753}
\definecolor{teal}{rgb}{0, 0.502, 0.502}

\makeatletter
\newdimen\itex@wd%
\newdimen\itex@dp%
\newdimen\itex@thd%
\def\itexspace#1#2#3{\itex@wd=#3em%
\itex@wd=0.1\itex@wd%
\itex@dp=#2ex%
\itex@dp=0.1\itex@dp%
\itex@thd=#1ex%
\itex@thd=0.1\itex@thd%
\advance\itex@thd\the\itex@dp%
\makebox[\the\itex@wd]{\rule[-\the\itex@dp]{0cm}{\the\itex@thd}}}
\makeatother

\makeatletter
\newif\if@sup
\newtoks\@sups
\def\append@sup#1{\edef\act{\noexpand\@sups={\the\@sups #1}}\act}%
\def\reset@sup{\@supfalse\@sups={}}%
\def\mk@scripts#1#2{\if #2/ \if@sup ^{\the\@sups}\fi \else%
  \ifx #1_ \if@sup ^{\the\@sups}\reset@sup \fi {}_{#2}%
  \else \append@sup#2 \@suptrue \fi%
  \expandafter\mk@scripts\fi}
\def\tensor#1#2{\reset@sup#1\mk@scripts#2_/}
\def\multiscripts#1#2#3{\reset@sup{}\mk@scripts#1_/#2%
  \reset@sup\mk@scripts#3_/}
\makeatother

\makeatletter
\newbox\slashbox \setbox\slashbox=\hbox{$/$}
\def\itex@pslash#1{\setbox\@tempboxa=\hbox{$#1$}
  \@tempdima=0.5\wd\slashbox \advance\@tempdima 0.5\wd\@tempboxa
  \copy\slashbox \kern-\@tempdima \box\@tempboxa}
\def\slash{\protect\itex@pslash}
\makeatother

\def\clap#1{\hbox to 0pt{\hss#1\hss}}

\let\oldroot\root
\def\root#1#2{\oldroot #1 \of{#2}}
\renewcommand{\sqrt}[2][]{\oldroot #1 \of{#2}}

\DeclareSymbolFont{symbolsC}{U}{txsyc}{m}{n}
\SetSymbolFont{symbolsC}{bold}{U}{txsyc}{bx}{n}
\DeclareFontSubstitution{U}{txsyc}{m}{n}

\DeclareSymbolFont{stmry}{U}{stmry}{m}{n}
\SetSymbolFont{stmry}{bold}{U}{stmry}{b}{n}

\DeclareFontFamily{OMX}{MnSymbolE}{}
\DeclareSymbolFont{mnomx}{OMX}{MnSymbolE}{m}{n}
\SetSymbolFont{mnomx}{bold}{OMX}{MnSymbolE}{b}{n}
\DeclareFontShape{OMX}{MnSymbolE}{m}{n}{
    <-6>  MnSymbolE5
   <6-7>  MnSymbolE6
   <7-8>  MnSymbolE7
   <8-9>  MnSymbolE8
   <9-10> MnSymbolE9
  <10-12> MnSymbolE10
  <12->   MnSymbolE12}{}

\makeatletter
\def\re@DeclareMathSymbol#1#2#3#4{%
    \let#1=\undefined
    \DeclareMathSymbol{#1}{#2}{#3}{#4}}
\re@DeclareMathSymbol{\neArrow}{\mathrel}{symbolsC}{116}
\re@DeclareMathSymbol{\neArr}{\mathrel}{symbolsC}{116}
\re@DeclareMathSymbol{\seArrow}{\mathrel}{symbolsC}{117}
\re@DeclareMathSymbol{\seArr}{\mathrel}{symbolsC}{117}
\re@DeclareMathSymbol{\nwArrow}{\mathrel}{symbolsC}{118}
\re@DeclareMathSymbol{\nwArr}{\mathrel}{symbolsC}{118}
\re@DeclareMathSymbol{\swArrow}{\mathrel}{symbolsC}{119}
\re@DeclareMathSymbol{\swArr}{\mathrel}{symbolsC}{119}
\re@DeclareMathSymbol{\nequiv}{\mathrel}{symbolsC}{46}
\re@DeclareMathSymbol{\Perp}{\mathrel}{symbolsC}{121}
\re@DeclareMathSymbol{\Vbar}{\mathrel}{symbolsC}{121}
\re@DeclareMathSymbol{\sslash}{\mathrel}{stmry}{12}
\re@DeclareMathSymbol{\boxslash}{\mathrel}{stmry}{27}
\re@DeclareMathSymbol{\boxbslash}{\mathrel}{stmry}{28}
\re@DeclareMathSymbol{\boxast}{\mathrel}{stmry}{24}
\re@DeclareMathSymbol{\boxcircle}{\mathrel}{stmry}{29}
\re@DeclareMathSymbol{\boxbox}{\mathrel}{stmry}{30}
\re@DeclareMathSymbol{\obslash}{\mathrel}{stmry}{20}
\re@DeclareMathSymbol{\obar}{\mathrel}{stmry}{58}
\re@DeclareMathSymbol{\olessthan}{\mathrel}{stmry}{60}
\re@DeclareMathSymbol{\ogreaterthan}{\mathrel}{stmry}{61}
\re@DeclareMathSymbol{\bigsqcap}{\mathop}{stmry}{"64}
\re@DeclareMathSymbol{\biginterleave}{\mathop}{stmry}{"6}
\re@DeclareMathSymbol{\invamp}{\mathrel}{symbolsC}{77}
\re@DeclareMathSymbol{\parr}{\mathrel}{symbolsC}{77}
\makeatother

\makeatletter
\def\Decl@Mn@Delim#1#2#3#4{%
  \if\relax\noexpand#1%
    \let#1\undefined
  \fi
  \DeclareMathDelimiter{#1}{#2}{#3}{#4}{#3}{#4}}
\def\Decl@Mn@Open#1#2#3{\Decl@Mn@Delim{#1}{\mathopen}{#2}{#3}}
\def\Decl@Mn@Close#1#2#3{\Decl@Mn@Delim{#1}{\mathclose}{#2}{#3}}
\Decl@Mn@Open{\llangle}{mnomx}{'164}
\Decl@Mn@Close{\rrangle}{mnomx}{'171}
\Decl@Mn@Open{\lmoustache}{mnomx}{'245}
\Decl@Mn@Close{\rmoustache}{mnomx}{'244}
\Decl@Mn@Open{\llbracket}{stmry}{'112}
\Decl@Mn@Close{\rrbracket}{stmry}{'113}
\makeatother

\makeatletter
\DeclareRobustCommand\widecheck[1]{{\mathpalette\@widecheck{#1}}}
\def\@widecheck#1#2{%
    \setbox\z@\hbox{\m@th$#1#2$}%
    \setbox\tw@\hbox{\m@th$#1%
       \widehat{%
          \vrule\@width\z@\@height\ht\z@
          \vrule\@height\z@\@width\wd\z@}$}%
    \dp\tw@-\ht\z@
    \@tempdima\ht\z@ \advance\@tempdima2\ht\tw@ \divide\@tempdima\thr@@
    \setbox\tw@\hbox{%
       \raise\@tempdima\hbox{\scalebox{1}[-1]{\lower\@tempdima\box
\tw@}}}%
    {\ooalign{\box\tw@ \cr \box\z@}}}
\makeatother

\makeatletter
\NewDocumentCommand\mathraisebox{moom}{%
\IfNoValueTF{#2}{\def\@temp##1##2{\raisebox{#1}{$\m@th##1##2$}}}{%
\IfNoValueTF{#3}{\def\@temp##1##2{\raisebox{#1}[#2]{$\m@th##1##2$}}%
}{\def\@temp##1##2{\raisebox{#1}[#2][#3]{$\m@th##1##2$}}}}%
\mathpalette\@temp{#4}}
\makeatletter

\makeatletter
\def\udots{\mathinner{\mkern2mu\raise\p@\hbox{.}
\mkern2mu\raise4\p@\hbox{.}\mkern1mu
\raise7\p@\vbox{\kern7\p@\hbox{.}}\mkern1mu}}
\makeatother

\newcommand{\Del}{\nabla}

\theoremstyle{plain}

\theoremstyle{definition}

\theoremstyle{remark}

\usetikzlibrary{positioning,positioning,arrows,decorations.markings}

\begin{document}
\renewcommand{\arraystretch}{1.5}

\preprint{
UTWI--30--2024\\
}

\title{On Twisted $A_{2n}$ Class-S Theories}

\author{Jacques Distler, Grant Elliot
     \oneaddress{
      Weinberg Institute for Theoretical Physics\\
      Department of Physics,\\
      University of Texas at Austin,\\
      Austin, TX 78712, USA \\
      {~}\\
      \email{distler@golem.ph.utexas.edu}\\
      \email{gelliot123@utexas.edu}
      }
}
\date{November 26, 2024}

\Abstract{In this paper, we investigate the twisted $A_{2n}$ sector of class-S theories. Heretofore, the Coulomb branches of such theories have been poorly understood. In this, and a companion paper, we  make progress in our understanding of them. In particular, we find a formula for the dimension of the Coulomb branch of any twisted $A_{2n}$ class-S theory. Deferring a systematic analysis to the companion paper, we here determine many contributions of punctures to the graded Coulomb branch dimensions, and in some low rank cases, all of them. We are then able to identify a variety of known 4d $\mathcal{N}=2$ SCFTs with twisted $A_{2n}$ theories, and reproduce many of their known properties, such as S-duality amongst certain Argyres-Douglas theories.}

\maketitle

\tocloftpagestyle{empty}
\tableofcontents
\vfill
\newpage
\setcounter{page}{1}

\section{Introduction}\label{introduction}
Four-dimensional $\mathcal{N}=2$ SCFTs have gained increasing attention over the past few decades. The existence of theories without a Lagrangian description can make understanding such theories difficult. Class-S theories \cite{Gaiotto:2009we,Gaiotto:2009hg} are somewhat well understood amongst non-Lagrangian theories, and are engineered by compactifying a 6d $\mathcal{N} =(2,0)$ theory on a Riemann surface. There are codimension-two defects labeled by nilpotent orbits\cite{Chacaltana:2012zy}, which can be placed at points along the Riemann surface, modifying the 4d theory. As the most general theory can be formed by gluing together Riemann surfaces, determining the set of theories corresponding to three punctured spheres is a very important problem. These have all been classified in a series of papers \cite{Chacaltana:2010ks,Chacaltana:2011ze,Chacaltana:2012ch,Chacaltana:2013oka,Chacaltana:2014jba,Chacaltana:2015bna,Chacaltana:2016shw,Chacaltana:2017boe,Chacaltana:2018vhp,Distler:2021cwz}, except for one case, the twisted $A_{2n}$ theories. There are various difficulties compared to other theories of class-S, as pointed out in \cite{Tachikawa:2011ch}. An infinite family of twisted $A_{2n}$ theories was studied in \cite{Chacaltana:2014nya} and some progress was made in the twisted $A_2$ case in \cite{Beem:2020pry} (see also \cite{Buican:2017fiq}).

Historically, the main issue with these theories is a lack of understanding of the Hitchin system that describes their Coulomb branch (CB). However, many of their other properties are known, such as the TQFT that describes their superconformal indices\cite{Gadde:2011uv,Mekareeya:2012tn} as well as their 3d mirrors \cite{Benini:2010uu,Beratto:2020wmn,Kang:2022zsl}. Moreover, in \cite{Beem:2022mde} the associated VOAs\cite{Beem:2013sza} were constructed, which in \cite{Elliot:2024hat} were shown to be related to the non-simply laced class-S VOAs of \cite{Arakawa:2018egx}. It was noted in \cite{Tachikawa:2018rgw} that the $Sp(n)$ flavour symmetry for the full puncture has a nontrivial global anomaly. Using a relation between the global anomaly and nilpotent Higgsing, we are able to determine the contribution of many punctures to the graded Coulomb branch dimensions. This leads us to identify an order reversing map of the set of $C_{n}$ nilpotent orbits, which determines the residue of the Higgs field at a twisted puncture. The analysis of the resulting Hitchin system is somewhat involved, and we defer it to an upcoming companion publication \cite{DistlerElliotWIP}.

In lieu of reading them off from the Hitchin system, in this paper we obtain the  graded Coulomb branch dimensions by studying the Higgs-branch RG flows between theories. Following \cite{Beem:2020pry}, the essential idea is that the contribution of a puncture to the graded Coulomb branch dimensions is a local phenomenon. That is, if we take two different theories with the same puncture $\mathcal{O}$, and then replace that puncture with $\mathcal{O}'$, the change in graded Coulomb branch dimensions is the same. Thus, we can determine the change in graded Coulomb branches from puncture replacement if the know how the corresponding Higgsing changes the graded Coulomb branch dimensions in a specific example. Hence, if the stratification of the Higgs branch of a twisted $A_{2n}$ class-S theory is known through other means, then the change in graded Coulomb branches for a puncture replacement can be determined by analyzing the corresponding Higgsing. 

We discuss general aspects of the Coulomb branches of twisted $A_{2n}$ theories in Section \ref{CB}. In Sections \ref{A4} and \ref{A6}, we use a variety of methods to determine the contributions of punctures to the graded Coulomb branch dimensions and catalog some of the ``interesting" theories in the twisted $A_4$ and $A_6$ case. In Section \ref{higher}, we briefly outline a similar endeavor for higher rank cases. We can perform a slew of additional consistency checks for the proposed graded Coulomb branch dimensions, such as compatibility with the Shapere-Tachikawa relation \cite{Shapere:2008zf}, S-duality, and the dimension of the Higgs branch of the 3d mirror.

\section{Coulomb Branch}\label{CB}

\subsection{Nilpotent Higgsing}
The particular RG flows we will make the most use of are called nilpotent Higgsings. At the level of the chiral algebra they act as Drinfeld-Sokolov(DS) reduction\cite{Beem:2014rza,Distler:2022nsn}. In any 4d $\mathcal{N}=2$ SCFT, a highest weight (HW) nilpotent Higgsing changes the anomaly coefficient $n_v$ by $k-1$, where $k$ is the level of the simple Higgsed factor. From the Shapere-Tachikawa relation, we see two straightforward ways to satisfy this: removing a CB parameter of dimension $k/2$ or replacing a CB parameter of dimension $k-1$ by one of dimension $\frac{k-1}{2}$. In class-S theories of other types, a highest weight nilpotent Higgsing has a consistent action on the graded Coulomb branch dimensions \cite{Distler:2022jvk}.  If the Higgsed simple factor has even level $k$, then a CB parameter with $\Delta=\frac{k}{2}$ is lost, whereas if $k$ is odd, a CB parameter with $\Delta= {k-1}$ is replaced by one with $\Delta =\frac{k-1}{2}$. The odd level case has to do with the fact that both Nahm orbits lie in the same special piece \cite{Distler:2022jvk}. This is not the case in the twisted $A_{2n}$ theories as can be seen looking at the examples in \cite{Beem:2020pry}. However, in the twisted $A_{2n}$ theory, a highest weight nilpotent Higgsing of a simple $Sp(n)$ factor from a twisted puncture of level $k$ has the following effect: if $k$ is odd a CB parameter with $\Delta=\frac{k}{2}$ is lost, whereas if $k$ is even, a CB parameter with $\Delta= {k-1}$ is replaced by one with $\Delta =\frac{k-1}{2}$. On the other hand, a HW nilpotent Higgsing of a manifest $SO(n)$ symmetry factor behaves the same as in the ordinary class-S case.

This reversed behavior for the $Sp(n)$ symmetry factors is due to the presence of Witten's global anomaly for the symmetry associated to a full puncture demonstrated in \cite{Tachikawa:2018rgw}, which we now explain. Recall the notion of an extended Coulomb branch (see \cite{Giacomelli:2020jel,Beem:2024fom} for more detailed discussions). A 4d $\mathcal{N}=2$ SCFT is said to have an extended Coulomb branch (ECB) if at a generic point in the Coulomb branch moduli space one finds free hypermultiplets in addition to the usual free vector multiplets.  From the point of view of the whole moduli space, one views the extended Coulomb branch as 
\begin{equation*}
    \frac{\widetilde{\text{CB}} \times \mathbb{H}^n}{\Gamma}
\end{equation*}
 where $\Gamma$ is a finite group and $\widetilde{\text{CB}}$ is a $|\Gamma|$ fold cover of the Coulomb branch of the theory. The intersection of the extended Coulomb branch  with the Higgs branch is $W = \frac{\mathbb{H}^n}{\Gamma}$\footnote{There are some subtleties that occur when the corresponding elementary slice of the Higgs branch is non-normal. Thankfully this is not the case for any of the examples in this paper.}. Giving a VEV living in $W$ results in a rank-preserving Higgsing. The Coulomb branch of the IR SCFT is then a cover of the Coulomb branch of the UV SCFT, as the above quotient space has the structure of a fiber bundle over $\mathbb{H}^n/\Gamma$ with fibers $\widetilde{\text{CB}}$.

 Suppose a theory has an $Sp(n)$ flavour symmetry that carries Witten's global anomaly. Going to a generic point of its Coulomb branch, we see that there must be some free hypermultiplets in order to match the global anomaly of the UV theory\footnote{Note that the global anomaly cannot be matched by a topological field theory from the arguments of \cite{Garcia-Etxebarria:2017crf}.}. Thus, the theory necessarily has an extended Coulomb branch. In particular, suppose we can do a highest weight nilpotent Higgsing of the simple factor containing the $Sp(n)$ symmetry. This stratum must then be contained in $W$, and is specifically $\mathbb{H}^n/\mathbb{Z}_2$. This must be a rank-preserving nilpotent Higgsing, and the resulting Coulomb branch should be a double cover of the Coulomb branch of the UV theory. From the Shapere-Tachikawa relation we see the only\footnote{We are assuming the Coulomb branch of the UV theory is freely generated.} way to satisfy the change in $n_v$ is for a CB parameter of dimension $k-1$ to be replaced by one of dimension $\frac{k-1}{2}$. Note that this cannot occur for $G \neq Sp(n)$ factors as the dimension of the fundamental representation does not equal the dimension of the minimal nilpotent orbit.

Conversely, if an $Sp(n)$ symmetry factor admits a HW nilpotent Higgsing but does not carry Witten's global anomaly, the flavour symmetry cannot act on the ECB hypers because there would be a violation of anomaly matching. Hence, the corresponding Higgsing must be rank-decreasing. We will see from our proposal for the 3d mirrors that the change in the rank must be one. Regardless, we will try to supplement evidence for this behavior of twisted $A_{2n}$ punctures with examples.

\subsection{Dimension of the Coulomb Branch}

For untwisted $A_{2n}$ class-S theories, the boundary conditions of $\mathcal{N}=4$ SYM play an important role \cite{Chacaltana:2012zy}. The action of S-duality on the boundary conditions is related to the singularity of the Higgs field of the Hitchin systems of corresponding class-S theories. There is a Nahm orbit specifying the boundary condition for the $\mathfrak{g}$ theory and a Hitchin orbit specifying the boundary condition for the S-dual $\mathfrak{g}^{\vee}$ theory. However, the compactification of the $A_{2n}$ $(2,0)$ theory with a twist line on a torus is somewhat mysterious, though as we will see we expect both the Nahm and Hitchin orbits in this case to be given by $C_n$ orbits. 
In particular, we expect there to be an order-reversing map on the set of $C_n$ orbits. An obvious guess is that this is the (original) Spaltenstein map \cite{Spaltenstein}, $d_S:\text{Nilp}(\mathfrak{g})\to \text{Nilp}(\mathfrak{g})$. That turns out not to be correct.

The 3d mirrors offer a clue into what the Hitchin orbits should be. For class-S theories, the 3d mirrors are star-shaped quivers with $T_{\rho}(G)$ theories attached to the central node, where $\rho$ indicates the puncture. For other $G$, the $T_{\rho}(G)$ theories have a Higgs branch equal to a (possibly trivial) cover\footnote{Depending on how one defines the $T_{\rho}(G)$ theories, they may always have their Higgs branch be exactly $d(\rho)$. The difference is due to a discrete gauging, and can often be seen in the UV quiver description via choices of orthogonal versus special orthogonal group factors.} of a certain nilpotent orbit of $\mathfrak{g}^{\vee}$, namely $d(\rho)$, where $d:\text{Nilp}(\mathfrak{g}^\vee)\to \text{Nilp}(\mathfrak{g})$ is the Spaltenstein(-Barbasch-Vogan) map \cite{BarbaschVogan}. This same orbit corresponds to the Hitchin orbit of the Hitchin system describing the Coulomb branch of the 4d theory. The Coulomb branch of the quiver is equal to the nilpotent Slodowy slice of $\rho$. However, in the twisted $A_{2n}$ case, the quivers attached to the central node for twisted punctures correspond to $T_{\rho}(Sp(n)')$ theories. Notably, $T_{[1^{2n}]}(Sp(n)')$ is self-mirror, and both its Coulomb branch and Higgs branch are equal to the nilpotent cone of $\mathfrak{sp}(n)$. We propose the Higgs branch of $T_{\rho}(Sp(n)')$ is the Hitchin orbit of the puncture $\rho$. Thus, we see the Hitchin orbit for the full puncture is equal to the regular orbit. Additionally, $T_{[2n]}(Sp(n)')$ turns out to just be 2n half-hypers which has Higgs branch equal to $\mathbb{C}^{2n}$. This leads us to assume that the Hitchin orbit for the regular Nahm orbit is the minimal nilpotent orbit, as the minimal nilpotent orbit of $Sp(n)$ is $\mathbb{C}^{2n}/\mathbb{Z}_2$. 

In \cite{Cremonesi:2014uva}, the Hilbert series of Higgs branches of various $T_{\rho}(Sp(n)')$ theories were computed, and the corresponding nilpotent orbits of $C_n$ were then deduced. We list their results in Table \ref{3dHiggs}.
\begin{table}
    \centering
    \begin{tabular}{|c|c|c|} \hline 
        $n$ & $\rho$ in $T_{\rho}(Sp(n)')$   & Higgs Branch \\ \hline 
        2 & $[2,1^2]$ & $[4]$\\
         & $[2^2]$ & $[2^2]$\\
        \hline
        3 & $[2,1^4]$ & $[6]$\\ 
        & $[2^2,1^2]$ & $[4,2]$\\ 
         &  $[2^3]$& $[4,2]$\\ 
        &  $[3^2]$& $[2^3]$\\
 & $[4,2]$&$[2^3]$\\ \hline
    \end{tabular}
    \caption{Higgs Branches determined by \cite{Cremonesi:2014uva}}
    \label{3dHiggs}
\end{table}

As discussed earlier, there should be an order reversing map on the set of nilpotent orbits of $C_n$, which we denote $d'$, that is different from the Spaltenstein map $d_S$. This map should describe the Higgs branch of the $T_{\rho}(Sp(n)')$. Much like the Spaltenstein map, $d'$ must satisfy $d'^3=d'$. To see this, consider the theory $T_{\rho}^{d'(\rho)}$. This theory has trivial Higgs branch, hence any additional Higgsing from $\rho$ to $d'^2(\rho)$ cannot lift the Higgs branch as there is nothing to lift. Therefore, $T_{\rho}(Sp(n)')$ and $T_{d'^2(\rho)}(Sp(n)')$ must have the same Higgs branch, and thus $d'=d'^3$.\footnote{This same argument implies there should be a similar order reversing map for any 3d $\mathcal{N}=4$ SCFT.} From what we have discussed so far, we can deduce a number of this map's properties.
\begin{enumerate}
    \item $d'([1^{2n}]) = d'([2,1^{2n-2}])= [2n]$
    \item $d'([2n]) = [2,1^{2n-2}]$
    \item  If there exists a nilpotent Higgsing of an even level $Sp(n)$ from $\mathcal{O}$ to $\mathcal{O}'$, then $d'(\mathcal{O})= d'(\mathcal{O}')$.
    \item If there exists a nilpotent Higgsing of an odd level $Sp(n)$ or any $SO(n)$ from $\mathcal{O}$ to $\mathcal{O}'$, then $\dim d'(\mathcal{O})=1+ \dim d'(\mathcal{O}')$
    \item $d' = d'^3$
\end{enumerate}
We can use these properties to determine its value in certain instances. For example, we know $[2^3] < d'([4,1^2]) < [4,2]$. Property 5 then determines that $d'([4,1^2]) = [4,1^2]$.

First, we must discuss what partitions are special\footnote{``Special" here refers to orbits in the image of this order reversing map. Additionally, a special piece refers to a maximal subset of orbits that are mapped to the same special orbit under this map.} for this new order reversing map. We only expect special pieces to come from orbits that are related by highest weight nilpotent Higgsings of $Sp(n)$ factors of even level\footnote{In the other classical types, the same statement holds but for $Sp(n)$ factors at odd level.}. From our argument that $d'=d'^3$, we see that for any orbit $\mathcal{O}$, $d'^2(\mathcal{O})$ is the maximal orbit in the special piece of $\mathcal{O}$ and the only special one. Thus, we propose the set of special orbits to be $C$-partitions which satisfy the property that for every odd part $i$, there is an odd number of even parts greater than $i$.  This is equivalent to the transpose of the partition being a $D$-partition.

Luckily, a map with these required properties has already appeared in the mathematics literature in \cite{MR1404331}\footnote{This same map can also be described as a map from $C_n$ orbits to $O_n$ orbits back to $C_n$ orbits as shown in \cite{MR4659865}.}. Orbits in its image are called metaplectic special (or anti-special). It can be given the following algorithmic description\cite{MR3666056}: Append 1 to a partition, then take the transpose and $C$-collapse. This will not actually be a $C$-partition, so we must subtract one from the last part of the partition. We propose that this is the map that describes the Higgs branch of $T_{\rho}(Sp(n)')$ theories, in addition to the residue of the Higgs field at the twisted punctures. While we leave a complete discussion of the Hitchin systems and constraints for a future publication\cite{DistlerElliotWIP}, we mention that one does indeed obtain $a$-type constraints for the odd degree invariant polynomials. This results in the presence of fractional graded Coulomb branch dimensions, which was how they were conjectured to arise in \cite{Beem:2020pry}.

Here are some examples:
\begin{equation*}
    [6,5^2] \to [6,5^2,1] \to [4,3^4,1] \to [4,3^4]
\end{equation*}
and
\begin{equation*}
    [6,3^2,1^2] \to [6,3^2,1^3] \to [6,3^2,1^3] \to [6,3^2,1^2].
\end{equation*}

Another order reversing map on C-partitions was discussed in \cite{Cremonesi:2014uva} that was also supposed to describe the Higgs branch of the $T_{\rho}(Sp(n)')$ theories, but disagrees with the one we have proposed. For example, in $C_3$ our map takes $[6]$ to $[2,1^4]$, while theirs takes $[6]$ to $[2^2,1^2]$. Notably, this results in their map not satisfying $d^3=d$. Physically, we expect the regular orbit to be taken to the minimal nilpotent orbit, since the corresponding quiver is just some free hypermultiplets. This leads us to believe the map we proposed is the correct one. 

Since this map defines an involution on the set of orbits in its image, one can construct symmetric Hasse diagrams that map an orbit to its image under the symmetry. The Hasse diagram for $C_2$ is
\begin{equation*}
\scalebox{.75}{
\begin{tikzpicture}
\node (0) at (0,0) {$[2,1^2]$};
\node[below=1cm of   0](22) {$[2^2]$};
\node[below=1cm of   22] (4) {$[4]$};
\path (0) edge (22)
 (22) edge (4)

;
\end{tikzpicture}
}.
\end{equation*}
The Hasse diagram for $C_3$ is
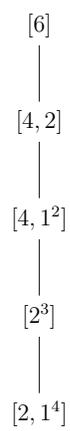
\begin{equation*}
\scalebox{.75}{
\begin{tikzpicture}
\node (0) at (0,0) {$[6]$};
\node[below=1cm of   0](22) {$[4,2]$};
\node[below=1cm of   22] (411) {$[4,1^2]$};
\node[below=1cm of   411] (222) {$[2^3]$};
\node[below=1cm of   222] (21111) {$[2,1^4]$};
\path (0) edge (22)
 (22) edge (411)
 (411) edge (222)
 (222) edge (21111)

;
\end{tikzpicture}
}.
\end{equation*}
The Hasse diagram for $C_4$ is
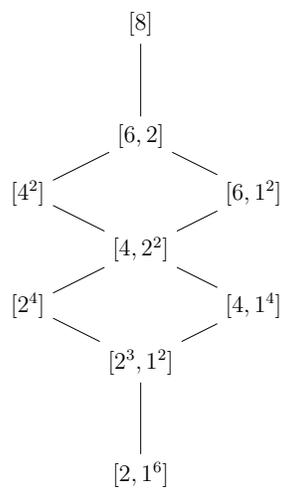
\begin{equation*}
\scalebox{.75}{
\begin{tikzpicture}
\node (0) at (0,0) {$[8]$};
\node at (0,-2) (62) {$[6,2]$};
\node at (2,-3) (611) {$[6,1^2]$};
\node at (-2,-3) (44) {$[4^2]$};
\node at (0,-4) (422) {$[4,2^2]$};
\node at (2,-5) (41111) {$[4,1^4]$};
\node at (-2,-5) (2222) {$[2^4]$};
\node at (0,-6) (22211) {$[2^3,1^2]$};
\node at (0,-8) (2111111) {$[2,1^6]$};
\path (0) edge (62)
 (62) edge (611)
 (62) edge (44)
 (44) edge (422)
 (611) edge (422)
 (422) edge (41111)
 (41111) edge (22211)
 (422) edge (2222)
 (2222) edge (22211)
 (22211) edge (2111111)

;
\end{tikzpicture}
}.
\end{equation*}
The Hasse diagram for $C_5$ is
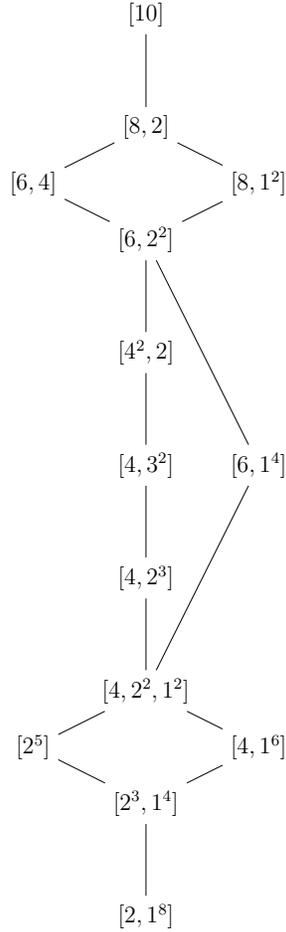
\begin{equation*}
\scalebox{.75}{
\begin{tikzpicture}
\node (0) at (0,0) {$[10]$};
\node at (0,-2) (82) {$[8,2]$};
\node at (2,-3) (811) {$[8,1^2]$};
\node at (-2,-3) (64) {$[6,4]$};
\node at (0,-4) (622) {$[6,2^2]$};
\node at (2,-8) (61111) {$[6,1^4]$};
\node at (0,-6) (442) {$[4^2,2]$};
\node at (0,-8) (433) {$[4,3^2]$};
\node at (0,-10) (4222) {$[4,2^3]$};
\node at (0,-12) (42211) {$[4,2^2,1^2]$};
\node at (-2,-13) (22222) {$[2^5]$};
\node at (2,-13) (4111111) {$[4,1^6]$};
\node at (0,-14) (2221111) {$[2^3,1^4]$};
\node at (0,-16) (211111111) {$[2,1^8]$};
\path (0) edge (82)
 (82) edge (811)
 (82) edge (64)
 (64) edge (622)
 (811) edge (622)
 (622) edge (61111)
 (61111) edge (42211)
 (622) edge (442)
 (442) edge (433)
 (433) edge (4222)
 (4222) edge (42211)
 (42211) edge (22222)
 (42211) edge (4111111)
 (22222) edge (2221111)
 (4111111) edge (2221111)
 (2221111) edge (211111111)

;
\end{tikzpicture}
}.
\end{equation*}

Denoting this map $d'$ and the Spaltenstein map $d$, we can then compute the dimension of a twisted $A_{2n}$ theory with untwisted punctures $\mathcal{O}_{1}....\mathcal{O}_n$ and twisted punctures $\mathcal{O}^{t}_1....\mathcal{O}^t_{m}$ as \begin{equation} \label{CBdim}
    \dim \text{Coulomb}=(g-1)\dim J+ \sum_{i}\dim d(\mathcal{O}_i)+\sum_{j} \Big(\dim d'(\mathcal{O}^t_j)+\frac{1}{2}(\dim J-\dim G^{\vee})\Big)
\end{equation}

\subsubsection{Highest Weight Nilpotent Higgsing}

Using the order reversing map, we can actually show the compatibility with nilpotent Higgsing and the predicted change in the dimension of the Coulomb branch. Let $\mathcal{O}_1$ and $\mathcal{O}_2$ be two $C_n$ orbits and let $\mathcal{O}_1'$ and $\mathcal{O}_2'$ be the two $C_{2n}$ orbits obtained by adding the part $2n$ to partitions $O_1$ and $\mathcal{O}_2$ respectively. Proposition 3.8 of \cite{MR4659865} can be used to show that if there is a Higgsing from $\mathcal{O}_1 \to \mathcal{O}_2$ in the twisted $A_{2n}$ theory, the change in the dimension of the Coulomb branch is the same as the change in dimension of the CB from the Higgsing $\mathcal{O}_{1}' \to \mathcal{O}_2'$ in the twisted $D_{2n+1}$ theory. To see this, note that a special case of their result gives the commutative diagram
\[
\begin{tikzcd}
  \{\mathcal{O} \in \text{Nilp}(\mathfrak{c}_{2n})| \mathbf{c}_1(\mathcal{O})=2n\} \arrow[r, "\hat{\Del}"] \arrow[d, "d"'] & \text{Nilp}(\mathfrak{c}_n) \arrow[d, "d'"] \\
  \{\hat{\mathcal{O}} \in \text{Nilp}^{\text{sp}}(\mathfrak{b}_{2n})|  \mathbf{r}_1(\hat{\mathcal{O}})=2n+1\} \arrow[r, "\Del"'] & \text{Nilp}(\mathfrak{c}_n)
\end{tikzcd}.
\] 
 $\hat{\Del}$ removes the the first column $2n$ from a partition, and $\Del$ removes the first row of a partition. A moment's reflection reveals that 
\begin{equation*}
    \dim(\hat{\mathcal{O}}_1)-\dim(\hat{\mathcal{O}}_2)=\dim(\Del(\hat{\mathcal{O}}_1))-\dim(\Del(\hat{\mathcal{O}}_2))
\end{equation*}
which combined with the commutative diagram gives the desired result
\begin{equation*}
    \dim(d(\mathcal{O}_1'))-\dim(d(\mathcal{O}_2'))=\dim(\Del(d(\mathcal{O}_1')))-\dim(\Del(d(\mathcal{O}_2')))=\dim(d'(\mathcal{O}_1))-\dim(d'(\mathcal{O}_2)).
\end{equation*}

\section{Twisted \texorpdfstring{$A_4$}{A₄}}\label{A4}

\subsection{Puncture Replacements}

\subsubsection{\texorpdfstring{$[1^4] \to [2,1^2]$}{[1⁴]→[2,1²]}}
We start with the $R_{2,4}$ theory initially investigated in \cite{Chacaltana:2014nya}. It has flavour symmetry $Sp(4)_6 \times U(1)$.  It is a rank-two theory with Coulomb branch operators of dimension 3 and 5. From the above, we expect a DS reduction of the manifest $Sp(2)_6$ to give two full hypermultiplets in addition to an interacting theory with $Sp(3)_5 \times U(1)$ flavour symmetry. The central charges of the interacting theory should be $a=61$ and $c=34$. There is a possible candidate theory in Table 6 of \cite{Martone:2021ixp} with graded Coulomb branch dimensions of $3$ and $\frac{5}{2}$. This theory was originally found in \cite{Zafrir:2016wkk} and later analyzed in \cite{Giacomelli:2020gee} and \cite{Martone:2021ixp}. This result is not surprising as it was argued in \cite{Giacomelli:2020gee} that this theory arises as a highest weight nilpotent Higgsing of $R_{2,4}$. Thus, we see an example of the reversed nilpotent Higgsing behavior.

Consider the rank-two theory discovered in \cite{Kaidi:2021tgr} which has a realization as a twisted $A_4$ theory with a wild puncture and a full regular puncture. Note when compactifying with a wild puncture the naive $U(1)_R$ symmetry mixes with the rotational symmetry of the cylinder, changing the flavour symmetry levels and the dimensions of Coulomb branch operators from that of the regular puncture. Nevertheless, the behavior under Higgsing remains the same. In this case the theory has an $Sp(2)_{\frac{13}{3}}$ symmetry and graded Coulomb branch parameters of dimensions $\frac{10}{3}$ and $\frac{4}{3}$. Performing a highest weight nilpotent Higgsing of the $Sp(2)$ gives the $(A_1,D_6)$ theory which has Coulomb branch parameters of dimensions $\frac{5}{3}$ and $\frac{4}{3}$. Evidently, a CB parameter of dimension $\frac{13}{3}-1=\frac{10}{3}$ was replaced by a dimension $\frac{5}{3}$ parameter. 

\subsubsection{\texorpdfstring{$[2,1^2] \to [2^2]$}{[2,1²]→[2²]}}

Going back to the $Sp(3)_5 \times U(1)$ theory and doing a DS reduction of the manifest $SU(2)_{5}$ should give 5 free hypermultiplets in addition to an interacting theory with $a=34$ and $c=19$. The theory should have flavour symmetry at least $Sp(2)_{4} \times U(1)$. The natural candidate is then the $C_2U_1$ theory which is expected from the Higgs Branch of the $Sp(3)_5\times U(1)$ theory as discussed in \cite{Martone:2021ixp}. The $C_2U_1$ theory has one Coulomb branch parameter of dimension $3$. We see a dimension $\frac{5}{2}$ operator has been removed as expected from nilpotent Higgsing which is consistent with the 3d mirror.

From the second example considered in the previous subsubsection, we see the $(A_1,D_6)$ AD theory has a realization as a twisted $A_4$ theory with a wild puncture and a $[2,1^2]$ regular puncture. Performing a highest weight nilpotent Higgsing of the $SU(2)_{\frac{10}{3}}$ results in the $(A_1,D_4)$ theory, hence we see a CB parameter of dimension $\frac{5}{3}$ was lost, as expected.

\subsubsection{\texorpdfstring{$[2^2] \to [4]$}{[2²]→[4]}}
Using the realization of the $Sp(2)_4 \times U(1)$ theory, we can't go further down the Hasse diagram of twisted punctures because we run into bad theories. However, we can go up the Hasse diagram along the untwisted puncture. Replacing the untwisted simple puncture by the $[3,2]$ puncture gives a rank-three theory with Coulomb branch parameters of dimensions $3,4$ and $5$. Now replacing the twisted puncture $[2^2]$ with $[4]$ gives the $D_2(SU(5))$ theory as argued in \cite{Beem:2020pry}. The $D_{p}(G)$ theories were introduced in \cite{Cecotti:2013lda,Cecotti:2012jx} This is a rank-two theory with graded Coulomb branch dimensions $\frac{3}{2}$ and $\frac{5}{2}$. Hence, the puncture replacement consists of removing a dimension 4 operator and replacing the dimension 3 and 5 operators with CB parameters of half their dimension. The rank of the theory decreased by one, as expected from the 3d mirror.

\subsubsection{Twisted Punctures}
\begin{tabular}{|c|c|c|c|c|c|}
\hline
\begin{tabular}{c}Nahm\\C-partition\end{tabular}&\begin{tabular}{c}Hitchin\\C-partition\end{tabular}&\begin{tabular}{c}Graded CB \\ Dimensions \\$\{2,3,4,5,\frac{3}{2},\frac{5}{2}\}$\end{tabular}&Flavour symmetry&$(\delta n_{h},\delta n_{v})$\\
\hline
$ [1^4]$&$[4]$&$\{1,\frac{5}{2},3,\frac{9}{2},0,0\}$&${Sp(2)}_6$&$(\frac{161}{2},77)$\\
\hline
$ [2,1^2]$&$[4]$&$\{1,\frac{5}{2},3,\frac{7}{2},0,1\}$&$SU(2)_5$&$(\frac{147}{2},72)$\\
\hline
$ [2^2]$&$[2^2]$&$\{1,\frac{5}{2},3,\frac{7}{2},0,0\}$&$U(1)$&$(\frac{137}{2},68)$\\
\hline
$[4]$&$[2,1^2]$&$\{1,\frac{3}{2},2,\frac{5}{2},1,1\}$&none&$(\frac{105}{2},53)$\\
\hline 
\end{tabular}

\subsection{Fixture Table}
We list fixtures with enhanced symmetries in Table \ref{TwistedA4}. We were able to determine the would-be unknown levels of various theories using Drinfeld-Sokolov reduction and S-duality. There is only one other rank-two theory other than the one mentioned earlier. It is fixture 16 in our table and has untwisted puncture $[3,2]$ and twisted punctures both given by $[2^2]$. It has Coulomb branch parameters of dimensions 3 and 4 and thus $n_v = 12$. The theory has $n_h= 18$ and calculating the Schur index one finds it has flavour symmetry $SU(2)^2 \times U(1)^2$. A similar enhancement occurs for fixture 15, and we can glue an irregular fixture to gauge the manifest $SU(2)_6$ which is S-dual to a $Sp(2)$ gauging of two $Sp(2)_4\times U(1)$ rank-one theories and 8 free hypers. This determines the levels of both $SU(2)$s to be 8 and the same must be true for fixture 16 since we could perform a Drinfeld-Sokolov reduction of the $SU(2)_6$. We find a matching theory in \cite{Martone:2021ixp} that also has flavour symmetry $SU(2)_8 \times SU(2)_8 \times U(1)^2$. This theory has also been studied in \cite{Martone:2021ixp, Zafrir:2016wkk, Giacomelli:2020gee}, and notably occurs at a strongly coupled point of the conformal manifold of $SU(4)$ gauge theory with a hypermultiplet in the symmetric and two in the fundamental.

\begin{longtable}{|c|c|c|c|c|}
\caption{Twisted $A_4$ Fixtures}\label{TwistedA4}\\
\hline
\#&Fixture& Flavour Symmetry&\begin{tabular}{c} Graded CB Dimensions \\ $\Delta_1,\Delta_2,...\Delta_r$\end{tabular}& ($n_h,n_v)$\\
\endfirsthead
\hline
\#&Fixture&Flavour Symmetry&Graded CB Dimensions& ($n_h,n_v)$\\
\endhead
\endfoot
\hline
1&$\begin{matrix} [1^4]\\ [1^4] \end{matrix}\quad [4,1]$&$\begin{gathered}{Sp(4)}_{6}\times U(1)\end{gathered}$&$\begin{gathered}3,5\end{gathered}$& (25,14)\\ 
\hline
2&$\begin{matrix} [2,1^2]\\ [1^4] \end{matrix}\quad [4,1]$&$\begin{gathered}{Sp(3)}_{5}\times U(1)+2 \end{gathered}$&$\begin{gathered}3,\frac{5}{2}\end{gathered}$& (18,9)\\ 
\hline
3&$\begin{matrix} [2^2]\\ [1^4] \end{matrix}\quad [4,1]$&$\begin{gathered}{Sp(2)}_{4}\times U(1)+4 \end{gathered}$&$\begin{gathered}3\end{gathered}$& (13,5)\\ 
\hline
4&$\begin{matrix} [2^2]\\ [1^4] \end{matrix}\quad [3,2]$&$\begin{gathered}Sp(2)_6 \times U(1)^2 \end{gathered}$&$\begin{gathered}3,4,5\end{gathered}$& (30,21)\\
\hline
5&$\begin{matrix} [4]\\ [2,1^2] \end{matrix}\quad [2,1^3]$&$\begin{gathered}{SU(2)}_5\times U(1) \times SU(3)_8\end{gathered}$&$\begin{gathered}\frac{3}{2},3,4,\frac{5}{2},\frac{5}{2}\end{gathered}$& (32,22)\\ 
\hline
6&$\begin{matrix} [4]\\ [2,1^2] \end{matrix}\quad [2^2,1]$&$\begin{gathered}{SU(2)}_5^3\times U(1)^2 \end{gathered}$&$\begin{gathered}\frac{3}{2},3,\frac{5}{2},\frac{5}{2}\end{gathered}$& (23,15)\\ 
\hline
7&$\begin{matrix} [4]\\ [2^2] \end{matrix}\quad [2,1^3]$&$\begin{gathered}{SU(2)}_{16} \times SU(3)_8 \times U(1)  \end{gathered}$&$\begin{gathered}\frac{3}{2},3,4,\frac{5}{2}\end{gathered}$& (27,18)\\ 
\hline
8&$\begin{matrix} [4]\\ [2^2] \end{matrix}\quad [2^2,1]$&$\begin{gathered}{SU(2)}_6 \times Sp(2)_5 \times U(1)\end{gathered}$&$\begin{gathered}\frac{3}{2},3,\frac{5}{2}\end{gathered}$& (18,11)\\ 
\hline
9&$\begin{matrix} [4]\\ [1^4] \end{matrix}\quad [3,1^2]$&$\begin{gathered}{Sp(2)}_5\times SU(2)_6\times U(1)+2 \end{gathered}$&$\begin{gathered}\frac{3}{2},3,\frac{5}{2}\end{gathered}$& (20,11)\\ 
\hline
10&$\begin{matrix} [4]\\ [1^4] \end{matrix}\quad [3,2]$&$\begin{gathered}{SU(5)}_5+2 \end{gathered}$&$\begin{gathered}\frac{3}{2},\frac{5}{2}\end{gathered}$& (14,6)\\ 
\hline
11&$\begin{matrix} [2,1^2]\\ [2,1^2] \end{matrix}\quad [3,1^2]$&$\begin{gathered}{SU(2)}_5^2 \times SU(2)_6 \times U(1) \end{gathered}$&$\begin{gathered}3,3,4,\frac{5}{2},\frac{5}{2}\end{gathered}$& (34,25)\\
\hline
12&$\begin{matrix} [2,1^2]\\ [2,1^2] \end{matrix}\quad [3,2]$&$\begin{gathered}{SU(2)}_5^2 \times U(1) \times U(1) \end{gathered}$&$\begin{gathered}3,4,\frac{5}{2},\frac{5}{2}\end{gathered}$& (28,20)\\
\hline
13&$\begin{matrix} [2^2]\\ [2,1^2] \end{matrix}\quad [3,1^2]$&$\begin{gathered}{SU(2)}_5 \times SU(2)_{16} \times SU(2)_6 \end{gathered}$&$\begin{gathered}3,3,4,\frac{5}{2}\end{gathered}$& (29,21)\\
\hline
14&$\begin{matrix} [2^2]\\ [2,1^2] \end{matrix}\quad [3,2]$&$\begin{gathered}{SU(2)}_5 \times SU(2)_{16} \times U(1) \end{gathered}$&$\begin{gathered}3,4,\frac{5}{2}\end{gathered}$& (23,16)\\
\hline
15&$\begin{matrix} [2^2]\\ [2^2] \end{matrix}\quad [3,1^2]$&$\begin{gathered}{SU(2)}_8 \times SU(2)_8 \times SU(2)_6 \end{gathered}$&$\begin{gathered}3,3,4\end{gathered}$& (24,17)\\ 
\hline
16&$\begin{matrix} [2^2]\\ [2^2] \end{matrix}\quad [3,2]$&$\begin{gathered}{SU(2)}_8 \times SU(2)_8 \times U(1)^2 \end{gathered}$&$\begin{gathered}3,4\end{gathered}$& (18,12)\\ 
\hline
17&$\begin{matrix} [4]\\ [4] \end{matrix}\quad [1^5]$&$\begin{gathered}{SU(5)}_5^2\end{gathered}$&$\begin{gathered}\frac{3}{2},\frac{3}{2},\frac{5}{2},\frac{5}{2}\end{gathered}$& (24,12)\\ 
\hline

\end{longtable}

Somewhat interesting is the fact that fixtures 8 and 9 appear to have the same interacting part. Indeed, computing the unrefined Schur index we find both to agree to at least order $\tau^{12}$. Following \cite{Bhardwaj:2021ojs,Distler:2022nsn} we can attempt to compute the global form of the flavour symmetry group for both fixtures, however the enhancements in fixture 8 make it impossible through these methods. Nevertheless, we can determine that both theories have $\gamma \gamma_1$ act trivially, where $\gamma =e^{2\pi i(j_1+j_2+R)}$ and $\gamma_1$ is the generator of the center of the $Sp(2)_5$.

This proposed isomorphism arises through a much different mechanism than has appeared in other class-S theories. In class-S theories of other type, such isomorphisms usually arise through some enhanced symmetry of a parent theory that ensures the RG flows obtained by turning on what appear to be different Higgs branch operators, are actually the same. In contrast, this instance involves a Higgsing of an $SU(2)$ from the parent theory and a sequence of Higgsings of an $Sp(2)$ and an $SU(2)$ which both lead to the same infrared theory. These are manifestly different RG flows and the isomorphism cannot be a result of some enhanced symmetry.

\begin{equation*}
    \begin{split}
   1+14 \tau ^2+16 \tau ^3+134 \tau ^4+224 \tau ^5+1050 \tau ^6+2032 \tau ^7+7030 \tau ^8+ 
   \\ 14336 \tau ^9+41384 \tau ^{10}+85488 \tau ^{11}+218760 \tau ^{12}+O\left(\tau ^{13}\right)
    \end{split}
\end{equation*}

We also note that this theory appears to have a realization as a generalized AD theory. In Table 7 of \cite{Li:2022njl} they list a theory that has the same invariants as ours. However, in that construction only the $U(1)_6 \subset SU(2)_6$ is manifest.

\subsection{Irregular Fixtures}

We note that any fixture with the two punctures $[4]$ and $[4,1]$ is a bad theory. Thus, we expect there to be an irregular fixture for the OPE of those two punctures. The only possible gauging comes from the $[2,1^2]$ puncture. We therefore need a a theory with an $SU(2)$ flavour symmetry at level three and no global anomaly. The only candidate is the $(A_1,D_4)$ Argyres-Douglas theory, also known as the rank-one $SU(3)$ instanton SCFT. One can check that this conjecture is consistent at the level of central charges and graded Coulomb branch dimensions.

\begin{longtable}{|c|c|c|c|c|}
\caption{Irregular $A_4$ Fixtures}\label{irregulara4}\\
\hline
\#&Fixture& Flavour Symmetry& Graded CB Dimensions& ($n_h,n_v)$\\
\endfirsthead
\hline
\#&Fixture&Old Flavour Symmetry&New Flavour Symmetry\\
\endhead
\endfoot
\hline
1&$\begin{matrix} [4]\\ [4,1] \end{matrix}\quad [([2,1^2], SU(2)]$&$\begin{gathered}SU(3)_3\end{gathered}$&$\begin{gathered}\frac{3}{2}\end{gathered}$& (4,2)\\ 
\hline
\end{longtable}

\subsection{AD Theories and S-duality}\label{ADSdual}

We note that fixture 6 appears to be isomorphic to what is called theory $T$ in \cite{Xie:2017vaf}. This matches up quite nicely with the fact that we can take fixture 17, which is two $D_2(SU(5))$ theories, glue an irregular fixture by adding a free hyper in the fundamental of an $SU(3)$, and then gauge. One can go to an alternative duality frame where this is theory $T$ coupled to a $D_2(SU(3))$ theory via an $SU(2)$ gauging. This same duality was first found in \cite{Xie:2016uqq}. That it is reproduced here serves as an excellent consistency check of our results. Pictorially, the first duality frame is given by
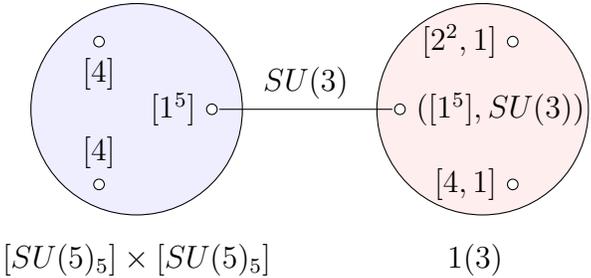
\begin{equation*}
\begin{tikzpicture}\draw[radius=40pt,fill=lightblue] circle;
\draw[radius=2pt,fill=white]  (-.5,.9) circle node[below=2pt] {$[4]$};
\draw[radius=2pt,fill=white]  (-.5,-1) circle node[above=2pt] {$[4]$};
\draw[radius=2pt,fill=white]  (1,0) circle node[left=2pt] {$[1^5]$};
\draw[radius=40pt,fill=lightred] (4.6,0) circle;
\draw[radius=2pt,fill=white]  (5,.9) circle node[left=2pt] {$[2^2,1]$};
\draw[radius=2pt,fill=white]  (5,-1) circle node[left=2pt] {$[4,1]$};
\draw[radius=2pt,fill=white]  (3.5,0) circle node[right=2pt] {$([1^5], SU(3))$};
\path
(1.1,0)  edge node[above] {$SU(3)$} (3.4,0);
\node at (0,-2) {$[SU(5)_5]\times [SU(5)_5]$};
\node at (4.5,-2) {$1(3)$};
\end{tikzpicture}
\end{equation*}
which is S-dual to
 
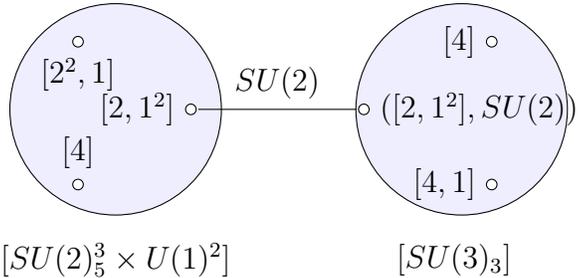
\begin{equation*}
\begin{tikzpicture}\draw[radius=40pt,fill=lightblue] circle;
\draw[radius=2pt,fill=white]  (-.5,.9) circle node[below=2pt] {$[2^2,1]$};
\draw[radius=2pt,fill=white]  (-.5,-1) circle node[above=2pt] {$[4]$};
\draw[radius=2pt,fill=white]  (1,0) circle node[left=2pt] {$[2,1^2]$};
\draw[radius=40pt,fill=lightblue] (4.6,0) circle;
\draw[radius=2pt,fill=white]  (5,.9) circle node[left=2pt] {$[4]$};
\draw[radius=2pt,fill=white]  (5,-1) circle node[left=2pt] {$[4,1]$};
\draw[radius=2pt,fill=white]  (3.3,0) circle node[right=2pt] {$([2,1^2], SU(2))$};
\path
(1.1,0)  edge node[above] {$SU(2)$} (3.2,0);
\node at (0,-2) {$[SU(2)_5^3 \times U(1)^2]$};
\node at (4.5,-2) {$[SU(3)_3]$};
\end{tikzpicture}.
\end{equation*}

\section{Twisted \texorpdfstring{$A_6$}{A₆}}\label{A6}

We will now perform a similar analysis of the twisted $A_6$ case. There are considerably more punctures to investigate, but our strategy remains largely the same. 

\subsection{Puncture Replacements}

\subsubsection{\texorpdfstring{$[1^6] \to [2,1^4] \to [2^2,1^2] \to [2^3]$}{[1⁶]→[2,1⁴]→[2²,1²]→[2³]}}
 We start at the fixture $[6,1],[1^6],[1^6]$. This has three CB generators of dimensions $\{3,5,7\}$ and flavour symmetry $Sp(6)_8 \times U(1)$. A nilpotent Higgsings/DS reduction of the $Sp(3)$ subalgebra associated to a twisted full puncture should take us to an interacting theory with flavour symmetry at least $Sp(5)_7 \times U(1)$ and central charges $(36,20)$ in addition to three hypermultiplets. A matching theory first appeared in  \cite{Zafrir:2016wkk} with CB generators of dimensions $\{3,5,\frac{7}{2}\}$. Thus, we see a dimension $7$ operator has been replaced by one of dimension $\frac{7}{2}$, as expected.

Doing another DS reduction of the $Sp(2)_7$ subalgebra gives the $R_{2,4}$ theory in addition to 6 free hypers. Thus, we see the puncture replacement $[2,1^4] \to [2^2,1^2]$ removes a dimension $\frac{7}{2}$ Coulomb branch operator. For the suspicious reader, we have computed their Schur indices to order $\tau^8$ and found both to be

\begin{equation}
    1+37 \tau ^2+713 \tau ^4+84 \tau ^5+9546 \tau ^6+2478 \tau ^7+99845 \tau ^8+O\left(\tau ^9\right).
\end{equation}

Doing another DS reduction of the $SU(2)_6$ gives the $Sp(3)_5\times U(1)$ rank-two theory. We can no longer do any more Higgsings without getting a bad theory. The identifications made so far agree exactly with the known behavior for highest weight nilpotent Higgsing.

\subsubsection{\texorpdfstring{$[2^3] \to [3^2]$}{[2³]→[3²]}}
Beginning with the above realization of the $Sp(3)_5 \times U(1)$ theory, we replace the untwisted $[6,1]$ puncture with the puncture $[5,2]$. The resulting fixture has graded Coulomb branch dimensions $3,4,(\frac{5}{2})^2,6$. The puncture replacement of $[2^3] \to [3^2]$ has $\delta n_v =15$. A plausible way for this to occur is losing a dimension 6 and $\frac{5}{2}$ CB operator. This would give us a rank-three theory with graded Coulomb branch dimensions $3,4,\frac{5}{2}$. We saw such a theory in the twisted $A_4$ case, specifically fixture 14 in Table \ref{TwistedA4}. The interacting parts of both fixtures have the same conformal anomalies and flavour symmetry given by $SU(2)_5 \times SU(2)_{16} \times U(1)$. Checking the global forms of their flavour symmetry, we find them to be the same. That is, the subgroup $\langle \gamma \gamma_1\gamma_3, \gamma_2 \rangle$ acts trivially, where $\gamma = e^{2 \pi i(j_1+j_2+R)}$ and $\gamma_1,\gamma_2$ and $\gamma_3$ are $\mathbb{Z}_2$ subgroups of the centers of $SU(2)_5, SU(2)_{16}$ and $U(1)$ respectively. Computing their Schur indices to order $\tau^8$ we find both to be

\begin{equation}
    1+7 \tau ^2+14 \tau ^3+57 \tau ^4+130 \tau ^5+410 \tau ^6+968\tau^7+2573\tau^8+O\left(\tau ^9\right).
\end{equation}

\subsubsection{\texorpdfstring{$[3^2] \to [4,2]$}{[3²]→[4,2]}}

In order to confirm the HW nilpotent Higgsings follows the predicted behavior, we turn our attention to the fixture with punctures $[5,2],[3^2]$ and $[1^6]$. This has manifest flavour symmetry $U(1) \times Sp(3)_8 \times SU(2)_8$. Calculating the Schur index to second order we find 12 moment map operators in the tensor product of the fundamentals of the two Sp algebras. Hence, there is an enhanced flavour symmetry $U(1) \times Sp(4)_8$. A DS reduction of the $Sp(3)_8$ takes us to a mixed fixture with graded CB dimensions $3,4,5,\frac{7}{2}$. Similarly, doing a DS reduction of the $SU(2)_8$ changes the punctures $[3^2] \to [4,2]$ and gives a mixed fixture with unknown CB dimensions. The interacting parts of these fixtures are the same, as they are both obtained by giving vevs to a Higgs branch operator in the same orbit of the symmetry, see \cite{Distler:2022kjb}. Hence, the nilpotent Higgsing of $[3^2] \to [4,2]$ also replaces a dimension $7$ operator with one of dimension $\frac{7}{2}$.

\subsubsection{\texorpdfstring{$[4,2] \to [6]$}{[4,2]→[6]}}

With the knowledge gathered so far, we can ascertain that the fixture $[3,4],[1^6],[4,2]$ has graded CB parameters of dimension $3,4,5,6,7,\frac{7}{2}$. The puncture replacement $[4,2] \to [6]$ should result in the $D_2(SU(7))$ theory in addition to three free hypermultiplets. This theory has graded Coulomb branch dimensions $\frac{3}{2},\frac{5}{2},\frac{7}{2}$. Thus, we know the puncture replacement removes CB parameters of dimensions 4, 6, and 7 while replacing dimension 3 and 5 parameters with dimension $\frac{3}{2}$ and $\frac{5}{2}$.

\subsubsection{\texorpdfstring{$[4,1^2] \to [4,2]$}{[4,1²]→[4,2]}}
There is still one twisted puncture $[4,1^2]$ whose contributions to the graded CB dimensions we have not determined. We can predict that Higgsing to $[4,2]$ should cause us to lose a dimension $\frac{5}{2}$ CB operator since the $SU(2)_5$ has trivial global anomaly, which implies going from $[2^3]$ to $[4,2]$ involves losing a dimension $6$ CB operator and a dimension 7 CB operator being replaced by one of dimension $\frac{7}{2}$. Unfortunately, we cannot find any alternative realizations of theories involving the $[4,1^2]$ puncture as additional confirmation. 

However, we can make an argument based on S-duality. Note that the collision of the punctures $[6]$ and $[6,1]$ should give an irregular puncture. After some thought, one realizes that the only possible gluing should be to the puncture $[4,1^2]$. We need a theory with an $SU(2)$ at level 3 that does not carry Witten's global anomaly. This is reminiscent of what occurred with the $[2,1^2]$ puncture in the $A_4$ theory, and the only candidate is again the rank-one $SU(3)$ instanton theory. Thus, knowing the contribution to the graded Coulomb branch dimensions of the punctures $[6,1]$ and $[6]$ we are able to confirm our prediction for $[4,1^2]$.

\subsection{Puncture Properties}

{
\renewcommand{\arraystretch}{1.5}
\begin{longtable}{|c|c|c|c|c|c|}
\hline
\begin{tabular}{c}Nahm\\C-partition\end{tabular}&\begin{tabular}{c}Hitchin\\C-partition\end{tabular}&\begin{tabular}{c}Graded CB \\ Dimensions \\$\{2,3,4,5,6,7,\frac{3}{2} ,\frac{5}{2}, \frac{7}{2}\}$\end{tabular}&\begin{tabular}{c}Flavour\\ symmetry\end{tabular}&$(\delta n_{h},\delta n_{v})$\\
\hline
\endhead 
$[1^6]$&$[6]$&$\{1,\frac{5}{2},3,\frac{9}{2},5,\frac{13}{2},0,0,0\}$&${Sp(3)}_8$&$(225,\tfrac{433}{2})$\\
\hline
$[2,1^4]$&$[6]$&$\{1,\frac{5}{2},3,\frac{9}{2},5,\frac{11}{2},0,0,1\}$&${Sp(2)}_7$&$(215,\tfrac{419}{2})$\\
\hline
$[2^2,1^2]$&$[4,2]$&$\{1,\frac{5}{2},3,\frac{9}{2},5,\frac{11}{2},0,0,0\}$&${SU(2)}_6\times U(1)$&$(207,\tfrac{407}{2})$\\
\hline
$[2^3]$&$[4,2]$&$\{1,\frac{5}{2},3,\frac{7}{2},5,\frac{11}{2},0,1,0\}$&$SU(2)_{24}$&$(201,\tfrac{397}{2})$\\
\hline
$[3^2]$&$[2^3]$&$\{1,\frac{5}{2},3,\frac{7}{2},4,\frac{11}{2},0,0,0\}$&$SU(2)_8$&$(185,\tfrac{367}{2})$\\
\hline
$[4,1^2]$&$[4,1^2]$&$\{1,\frac{5}{2},3,\frac{7}{2},4,\frac{9}{2},0,1,1\}$&${SU(2)}_5$&$(182,\tfrac{361}{2})$\\
\hline
$[4,2]$&$[2^3]$&$\{1,\frac{5}{2},3,\frac{7}{2},4,\frac{9}{2},0,0,1\}$&none&$(177,\tfrac{353}{2})$\\
\hline
$[6]$&$[2,1^4]$&$\{1,\frac{3}{2},2,\frac{5}{2},3,\frac{7}{2},1,1,1\}$&none&$(137,\tfrac{275}{2})$\\
\hline
\end{longtable}
}

\begin{longtable}{|c|c|c|c|c|}
\caption{Useful Twisted $A_6$ Fixtures}\label{TwistedA6}\\
\hline
\#&Fixture& Flavour Symmetry& Graded CB Dimensions& ($n_h,n_v)$\\
\endfirsthead
\hline
\#&Fixture&Old Flavour Symmetry&Graded CB Dimensions & $(n_h,n_v)$\\
\endhead
\endfoot
\hline
1&$\begin{matrix} [1^6]\\ [1^6] \end{matrix}\quad [6,1]$&$\begin{gathered}{Sp(6)}_{8}\times U(1)\end{gathered}$&$\begin{gathered}3,5,7\end{gathered}$& (49,27)\\ 
\hline
2&$\begin{matrix} [1^6]\\ [2,1^4] \end{matrix}\quad [6,1]$&$\begin{gathered}{Sp(5)}_{7}\times U(1)+3 \end{gathered}$&$\begin{gathered}3,5,\frac{7}{2}\end{gathered}$& (39,20)\\ 
\hline
3&$\begin{matrix} [1^6]\\ [2^2,1^2] \end{matrix}\quad [6,1]$&$\begin{gathered}{Sp(4)}_{6}\times U(1)+6 \end{gathered}$&$\begin{gathered}3,5\end{gathered}$& (31,14)\\ 
\hline
4&$\begin{matrix} [2^3]\\ [1^6] \end{matrix}\quad [6,1]$&$\begin{gathered}Sp(3)_5 \times U(1)+9 \end{gathered}$&$\begin{gathered}3,\frac{5}{2}\end{gathered}$& (27,9)\\
\hline
5&$\begin{matrix} [2^3]\\ [3^2] \end{matrix}\quad [5,2]$&$\begin{gathered}{SU(2)}_5\times SU(2)_{16}\times U(1)+2 \end{gathered}$&$\begin{gathered}3,4,\frac{5}{2}\end{gathered}$& (25,16)\\ 
\hline
6&$\begin{matrix} [1^6]\\ [3^2] \end{matrix}\quad [5,2]$&$\begin{gathered}Sp(4)_8 \times U(1) \end{gathered}$&$\begin{gathered}3,4,5,7\end{gathered}$& (50,34)\\ 
\hline
\end{longtable}

\subsection{Irregular Fixtures}

We can perform a similar analysis to what was done in the twisted $A_4$ case. We note that no free hypers can appear as irregular fixtures as there would be no anomaly-free gauging of the $Sp(n)$. These irregular fixtures give additional evidence for the proposed graded Coulomb branch dimensions.  

\begin{longtable}{|c|c|c|c|c|}
\caption{Irregular $A_6$ Fixtures}\label{IrregularA6}\\
\hline
\#&Fixture& Flavour Symmetry& Graded CB Dimensions& $(n_h,n_v)$\\
\endfirsthead
\hline
\#&Fixture&Flavour Symmetry&Graded CB dimensions&$(n_h,n_v)$\\
\endhead
\endfoot
\hline
1&$\begin{matrix} [6]\\ [6,1] \end{matrix}\quad \bigl([4,1^2], SU(2)\bigr)$&$\begin{gathered}SU(3)_3\end{gathered}$&$\begin{gathered}\frac{3}{2}\end{gathered}$& (4,2)\\ 
\hline
2&$\begin{matrix} [6]\\ [5,2] \end{matrix}\quad \bigl([2,1^4], Sp(2)\bigr)$&$\begin{gathered}SU(5)_5\end{gathered}$&$\begin{gathered}\frac{3}{2},\frac{5}{2}\end{gathered}$& (12,6)\\ 
\hline
3&$\begin{matrix} [6]\\ [5,1^2] \end{matrix}\quad \bigl([2,1^4], Sp(2)\bigr)$&$\begin{gathered}Sp(2)_5\times SU(2)_6 \times U(1)\end{gathered}$&$\begin{gathered}\frac{3}{2},3,\frac{5}{2}\end{gathered}$& (18,11)\\ 
\hline
4&$\begin{matrix} [3^2]\\ [6,1] \end{matrix}\quad \bigl([1^6], Sp(2)\bigr)$&$\begin{gathered}Sp(2)_4\times U(1)\end{gathered}$&$\begin{gathered}3\end{gathered}$& (9,5)\\ 
\hline
5&$\begin{matrix} [4,2]\\ [6,1] \end{matrix}\quad \bigl([2,1^4], Sp(2)\bigr)$&$\begin{gathered}Sp(2)_4\times U(1)+\frac{1}{2}(4)\end{gathered}$&$\begin{gathered}3\end{gathered}$& (9,5)\\ 
\hline
6&$\begin{matrix} [4,1^2]\\ [6,1] \end{matrix}\quad \bigl([2,1^4], Sp(2)\bigr)$&$\begin{gathered}Sp(3)_5\times U(1)\end{gathered}$&$\begin{gathered}3,\frac{5}{2}\end{gathered}$& (18,9)\\ 
\hline
\end{longtable}

\section{Higher Rank}\label{higher}

We conclude with some general remarks we can make on higher rank twisted $A_{2n}$ theories. The fixture $[2n,1],[1^{2n}],[2,1^{2n-2}]$ appears to be the $Sp(2n-1) \times U(1)$ theory identified in \cite{Zafrir:2016wkk}. We can perform additional highest weight nilpotent Higgsings of the $Sp(n)$ symmetry to obtain more theories in the $Sp(N) \times U(1)$ series.

We mention that all twisted $A_{2N}$ theories with $N>1$ have a puncture $[2N-2,1^2]$ which contributes a flavour symmetry $SU(2)_5$. There is an irregular fixture given by 
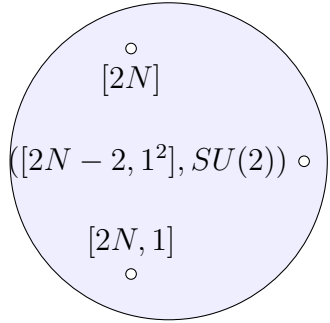
\begin{equation*}
\begin{tikzpicture}\draw[radius=60pt,fill=lightblue] circle;
\draw[radius=2pt,fill=white]  (-.5,1.5) circle node[below=2pt] {$[2N]$};
\draw[radius=2pt,fill=white]  (-.5,-1.5) circle node[above=2pt] {$[2N,1]$};
\draw[radius=2pt,fill=white]  (1.8,0) circle node[left=2pt] {$([2N-2,1^2],SU(2))$};
\end{tikzpicture}
\end{equation*}
which corresponds to the rank-one $SU(3)_3$ theory.
Now consider the realization of two copies of the $D_2(SU(2N+1))$ theory. We can tensor this with a free hypermultiplet in the fundamental of the diagonal $SU(2N+1)$ and then perform the exactly marginal gauging to obtain the four punctured sphere
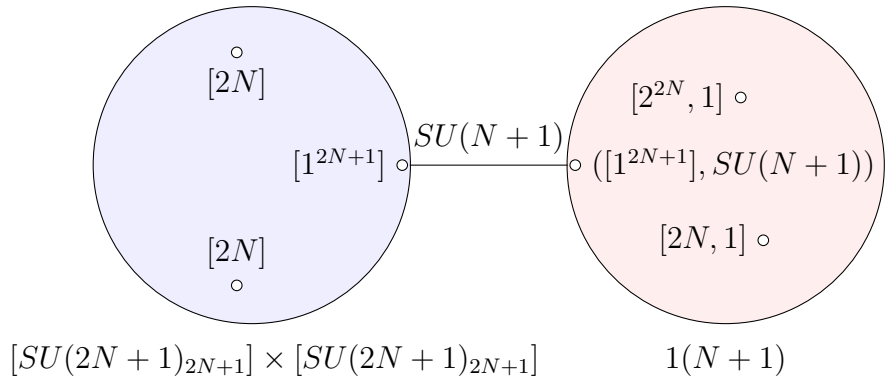
\begin{equation*}
\begin{tikzpicture}\draw[radius=60pt,fill=lightblue] (-.3,0) circle;
\draw[radius=2pt,fill=white]  (-.5,1.5) circle node[below=2pt] {$[2N]$};
\draw[radius=2pt,fill=white]  (-.5,-1.6) circle node[above=2pt] {$[2N]$};
\draw[radius=2pt,fill=white]  (1.7,0) circle node[left=2pt] {$[1^{2N+1}]$};
\draw[radius=60pt,fill=lightred] (6,0) circle;
\draw[radius=2pt,fill=white]  (6.2,.9) circle node[left=2pt] {$[2^{2N},1]$};
\draw[radius=2pt,fill=white]  (6.5,-1) circle node[left=2pt] {$[2N,1]$};
\draw[radius=2pt,fill=white]  (4,0) circle node[right=2pt] {$([1^{2N+1}], SU(N+1))$};
\path
(1.8,0)  edge node[above] {$SU(N+1)$} (3.9,0);
\node at (0,-2.6) {$[SU(2N+1)_{2N+1}]\times [SU(2N+1)_{2N+1}]$};
\node at (6,-2.6) {$1(N+1)$};
\end{tikzpicture}.
\end{equation*}
Going to the other S-duality frame gives
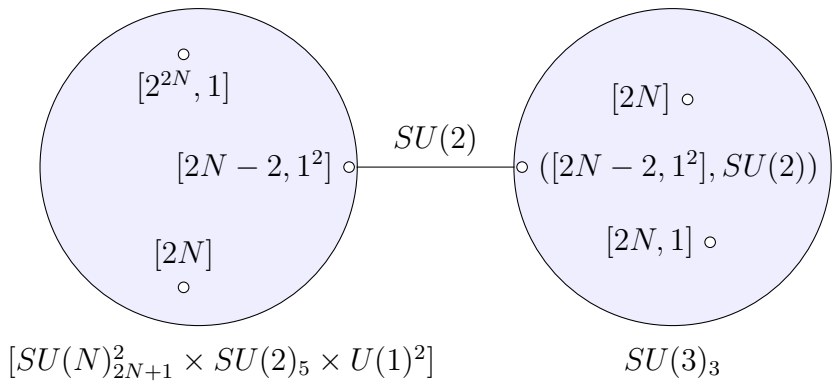
\begin{equation*}
\begin{tikzpicture}\draw[radius=60pt,fill=lightblue] (-.3,0) circle;
\draw[radius=2pt,fill=white]  (-.5,1.5) circle node[below=2pt] {$[2^{2N},1]$};
\draw[radius=2pt,fill=white]  (-.5,-1.6) circle node[above=2pt] {$[2N]$};
\draw[radius=2pt,fill=white]  (1.7,0) circle node[left=2pt] {$[2N-2,1^2]$};
\draw[radius=60pt,fill=lightblue] (6,0) circle;
\draw[radius=2pt,fill=white]  (6.2,.9) circle node[left=2pt] {$[2N]$};
\draw[radius=2pt,fill=white]  (6.5,-1) circle node[left=2pt] {$[2N,1]$};
\draw[radius=2pt,fill=white]  (4,0) circle node[right=2pt] {$([2N-2,1^2], SU(2))$};
\path
(1.8,0)  edge node[above] {$SU(2)$} (3.9,0);
\node at (0,-2.6) {$[SU(N)_{2N+1}^2 \times SU(2)_5 \times U(1)^2]$};
\node at (6,-2.6) {$SU(3)_3$};
\end{tikzpicture}.
\end{equation*}

The same S-duality frame of this gauging appears in \cite{Xie:2016uqq}. Evidently, the fixture on the left is identified with what they refer to as a class-B theory. In the language of \cite{Xie:2012hs}, it is a class-S theory engineered with a type III irregular puncture with data $Y_1=[1,1,...,1]$, $Y_2=[N-1,N-1,1,1]$ and $Y_3=[N-1,N-1,1,1]$. From Equation \ref{CBdim} we see that this theory has rank $3N-2$, which agrees with the duality. The case $N=2$ corresponds to the duality we discussed in Subsection \ref{ADSdual}.

More generally, in the twisted $A_{2N}$ theory the punctures $[1^{2k},2N-k]$ carry a $Sp(k)_{2k+3}$ which can be glued to a $D_{2}(SU(k+1))$ theory via a gauging of the diagonal $Sp(k)_{2k+4}$. The commutant of this is a $U(1)$, and so the corresponding irregular fixture should be obtained as the OPE of two punctures whose total contribution to the flavour symmetry is a $U(1)$. The untwisted puncture must contribute the $U(1)$, so it is of the form $[a,2N+1-a]$. From our examples we expect that the twisted puncture should be $[2N]$ and the untwisted one should be $[2N+1-k,k]$.

Using the above conjectured irregular fixtures and nilpotent Higgsing, we can solve for all the graded CB dimensions of punctures in the twisted $A_8$ theory.

{
\renewcommand{\arraystretch}{1.5}
\begin{longtable}{|c|c|c|c|c|c|}
\hline
\mbox{\shortstack{\\Nahm\\C-partition}}&\mbox{\shortstack{\\Hitchin\\C-partition}}&\begin{tabular}{c}Graded CB \\ Dimensions \\$\{2,3,4,5,6,7,8,9,\frac{3}{2} ,\frac{5}{2}, \frac{7}{2},\frac{9}{2}\}$\end{tabular}&Flavour symmetry&$(\delta n_{h},\delta n_{v})$\\
\hline 
\endhead
$[1^8]$&$[8]$&$\{1,\frac{5}{2},3,\frac{9}{2},5,\frac{13}{2},7,\frac{17}{2},0,0,0,0\}$&${Sp(4)}_{10}$&$(\frac{963}{2},466)$\\
\hline
$[2,1^6]$&$[8]$&$\{1,\frac{5}{2},3,\frac{9}{2},5,\frac{13}{2},7,\frac{15}{2},0,0,0,1\}$&${Sp(3)}_9$&$(\frac{937}{2},457)$\\
\hline
$[2^2,1^4]$&$[6,2]$&$\{1,\frac{5}{2},3,\frac{9}{2},5,\frac{13}{2},7,\frac{15}{2},0,0,0,0\}$&${Sp(2)}_8\times U(1)$&$(\frac{915}{2},449)$\\
\hline 
$[2^3,1^2]$&$[6,2]$&$\{1,\frac{5}{2},3,\frac{9}{2},5,\frac{11}{2},7,\frac{15}{2},0,0,1,0\}$&${SU(2)}_{32}\times {SU(2)}_7$&$(\frac{897}{2},442)$\\
\hline 
$[2^4]$&$[4^2]$&$\{1,\frac{5}{2},3,\frac{9}{2},5,\frac{11}{2},7,\frac{15}{2},0,0,0,0\}$&${SU(2)}_{16}^2$&$(\frac{883}{2},436)$\\
\hline 
$[3^2,1^2]$&$[4,2^{2}]$&$\{1,\frac{5}{2},3,\frac{9}{2},5,\frac{11}{2},6,\frac{15}{2},0,0,0,0\}$&${SU(2)}_{10}\times {SU(2)}_6$&$(\frac{851}{2},421)$\\
\hline 
$[4,1^4]$&$[6,1^2]$&$\{1,\frac{5}{2},3,\frac{9}{2},5,\frac{11}{2},6,\frac{13}{2},0,0,1,1\}$&${Sp(2)}_7$&$(\frac{847}{2},418)$\\
\hline 
$[4,2,1^2]$&$[4,2^2]$&$\{1,\frac{5}{2},3,\frac{9}{2},5,\frac{11}{2},6,\frac{13}{2},0,0,0,1\}$&${SU(2)}_6$&$(\frac{831}{2},412)$\\
\hline 
$[3^2,2]$&$[4,2^2]$&$\{1,\frac{5}{2},3,\frac{7}{2},5,\frac{11}{2},6,\frac{15}{2},0,1,0,0\}$&${SU(2)}_{10}$&$(\frac{839}{2},416)$\\
\hline 
$[4,2^2]$&$[4,2^2]$&$\{1,\frac{5}{2},3,\frac{7}{2},5,\frac{11}{2},6,\frac{13}{2},0,1,0,1\}$&$U(1)$&$(\frac{819}{2},407)$\\
\hline 
$[4^2]$&$[2^4]$&$\{1,\frac{5}{2},3,\frac{7}{2},4,\frac{11}{2},6,\frac{13}{2},0,0,0,0\}$&$U(1)$&$(\frac{771}{2},384)$\\
\hline 
$[6,1^2]$&$[4,1^4]$&$\{1,\frac{5}{2},3,\frac{7}{2},4,\frac{9}{2},5,\frac{11}{2},0,1,1,1\}$&${SU(2)}_5$&$(\frac{717}{2},357)$\\
\hline 
$[6,2]$&$[2^3,1^2]$&$\{1,\frac{5}{2},3,\frac{7}{2},4,\frac{9}{2},5,\frac{11}{2},0,0,1,1\}$&none&$(\frac{707}{2},353)$\\
\hline 
$[8]$&$[2,1^6]$&$\{1,\frac{3}{2},2,\frac{5}{2},3,\frac{7}{2},4,\frac{9}{2},1,1,1,1\}$&none&$(\frac{563}{2},282)$\\
\hline 
\end{longtable}
}

\section{Discussion}

In this paper, we have made some headway in understanding the graded Coulomb branch dimensions of twisted $A_{2n}$ theories in addition to their irregular fixtures. Moreover, we have determined the residue of the Higgs field at twisted punctures. In an upcoming publication \cite{DistlerElliotWIP}, we will discuss the Hitchin systems in detail along with the corresponding constraints. It would be interesting to know exactly what subset of these theories overlap with those of class-S theories obtained from compactification with a wild puncture. The regular class-S description may offer some insight to the VOAs of these generalized AD theories, at least in cases where they are not known yet.  

A particularly intriguing direction for future research would be to understand the connection between the order-reversing map of \cite{MR1404331} and the S-duality of boundary conditions in $\mathcal{N}=4$ SYM theory obtained by compactifying the $A_{2n} $theory on a torus with a twist line. What is the relation to the story in \cite{Gaiotto:2008ak}? We leave a full investigation for future work.

\section*{Acknowledgements}
 This work was supported in part by the National Science Foundation under Grant No.~PHY--2210562. 
\bibliographystyle{utphys}
\bibliography{references}

\end{document}